\documentclass[nonacm]{acmart}

\AtBeginDocument{%
  \providecommand\BibTeX{{%
    \normalfont B\kern-0.5em{\scshape i\kern-0.25em b}\kern-0.8em\TeX}}}

\usepackage{dirtytalk}
\usepackage{csquotes}
\begin{document}

\title{Bridging the Gap: \\ Integrating Ethics and Environmental Sustainability in  AI Research and Practice}

\author{Alexandra Sasha Luccioni}
\author{Giada Pistilli}
\email{sasha.luccioni@huggingface.co}
\affiliation{%
  \institution{Hugging Face}
  \country{Canada/France}
}

\author{Raesetje Sefala}
\author{Nyalleng Moorosi}
\affiliation{%
  \institution{Distributed AI Research Institute}
  \country{Canada/Lesotho}
}

\begin{abstract}
As the possibilities for Artificial Intelligence (AI) have grown, so have concerns regarding its impacts on society and the environment. However, these issues are often raised separately; i.e. carbon footprint analyses of AI models typically do not consider how the pursuit of scale has contributed towards building models that are both inaccessible to most researchers in terms of cost and disproportionately harmful to the environment. On the other hand, model audits that aim to evaluate model performance and disparate impacts mostly fail to engage with the environmental ramifications of AI models and how these fit into their auditing approaches. In this separation, both research directions fail to capture the depth of analysis that can be explored by considering the two in parallel and the potential solutions for making informed choices that can be developed at their convergence. In this essay, we build upon work carried out in AI and in sister communities, such as philosophy and sustainable development, to make more deliberate connections around topics such as generalizability, transparency,  evaluation and equity across AI research and practice. We argue that the efforts aiming to study AI's ethical ramifications should be made in tandem with those evaluating its impacts on the environment, and we conclude with a proposal of best practices to better integrate AI ethics and sustainability in AI research and practice.  
\end{abstract}

\maketitle

\section{Introduction} \label{sec:intro}

In recent years, AI systems have become pervasive, presented as a key tool in the fight against climate change~\cite{rolnick2022tackling} and in societally-beneficial domains such as health and education~\cite{floridi2018ai4people,ghassemi2020review,holmes2022state}. However, the training and deployment of AI systems also comes with a cost in terms of energy and natural resources~\cite{strubell2019energy,luccioni2023power, dobbe2019ai} and can inadvertently result in the amplification of inequalities~\cite{selbst2019fairness} and proliferation of biases~\cite{angwin2022machine} when systems are put into practice. Historically, the societal and environmental impacts of AI systems have been addressed separately, in two distinct areas of study -- i.e. scholarship that aims to address the ethics of AI models focuses on aspects such as bias evaluation or auditing~\cite{buolamwini2018gender,utama2020towards}, typically overlooking models' impacts on natural resources and ecosystems~\cite{mehrabi2021survey}. Conversely, work that aims to estimate the growing carbon footprint and energy consumption of AI models~\cite{strubell2019energy, luccioni2022estimating} does not typically address the contribution this has towards shifting the balance of power or amplifying inequalities~\cite{abdalla2023elephant,abdalla2021grey}. 

Some recent scholarship has started to establish explicit links between ethics and sustainability often hones in on specific applications, such as the emblematic "Stochastic Parrots" paper, which addresses both sustainability and ethics as issues in the context of large language models (LLMs)~\cite{bender2021dangers}. Apart from this exception and a precious few others, conversations around AI ethics and sustainability have taken place separately, in different venues and by different sets of stakeholders. However, since in both AI sustainability and AI ethics research, the aim is to think through how we might develop technologies which are useful and inclusive while we limit harm to people and environments, similar themes emerge, often structured via concepts such as functionality and efficiency just as much as justice, fairness and equity. In both of these disciplines. 

In the present article, we highlight these transversal themes and argue that both the societal and environmental impacts of AI systems should be considered in parallel in all aspects of AI theory, practice and governance. Drawing upon work from within the AI community as well as related fields, we explore a series of use cases illustrating that ethics and sustainability go hand-in-glove when it comes to the development and the deployment of AI systems. By doing so, we hope to shed light on the importance of pursuing research that blends together different perspectives to allow for a better understanding of the societal impacts of AI systems. 

Given that our goal is to allow our work to be read and understood by a variety of audiences, we start, in Section~\S\ref{sec:definition}, by defining the concepts and terms that are core to our subsequent analysis. We continue by examining the current state of AI ethics and sustainability from 3 different but complementary perspectives:  theoretical principles and frameworks (\S\ref{subsec:principles}), AI research and practice(\S\ref{subsec:research}), and AI regulation and governance (\S\ref{subsection:regulation}). Next, in \S\ref{sec:links}, we identify four transversal themes that we have found to be particularly central to both AI ethics and sustainability: generalizability, evaluation, transparency, and power. Finally, in \S\ref{sec:ways-forward}, we propose ways forward at the intersection of these themes and three directions of AI research and practice. We wrap up the article with ideas for future work that can be pursued at the nexus of AI ethics and sustainability and a brief conclusion.

\section{Key Concepts and Definitions} \label{sec:definition}

\subsection*{Sustainability}

When it comes to the concept of sustainability, one of its first and most widely-accepted definitions originates from the 1987 Brundtland Report, which defines sustainable development as \textit{“development that meets the needs of the present without compromising the ability of future generations to meet their own needs”}~\citeyearpar{imperatives1987report}[p.41]. This definition remains central to the field of sustainability write large, informing frameworks such as the UN Sustainable Development Goals (SDGs), which were developed in 2015 as a blueprint for achieving peace and prosperity for people and the planet~\cite{unsdgs2015}. However, also in 1987, environmental economist Edward Barbier proposed an alternative definition to sustainability, structuring it around three pillars: environmental, societal and economic, arguing that sustainable development can only be truly achieved when both environmental stewardship, social equity and economic viability coexist and are inter-connected~\citeyearpar{barbier1987concept}. 

In the context of AI, the term `sustainability' is most often used to refer solely to environmental sustainability~\cite{cowls2023ai,falk2023challenging}. The umbrella term `Sustainable AI' was initially proposed by van Wynsberghe as a field of practice that both aims to use AI in climate-positive applications, as well as improving upon the (environmental) sustainability of AI approaches themselves~\cite{van2021sustainable}. This proposal would then encompass the vast variety of work being done at the nexus of machine learning and fields such as biodiversity monitoring, agriculture, transportation, etc. (for a review, see~\cite{vinuesa2020role} and ~\cite{rolnick2022tackling}). AI and sustainability has also been central to workshops such as SustaiNLP and the International Sustainable AI Workshop, that have put the emphasis on efficient methods and the application of AI to sustainability-related problems, as well as the Tackling Climate Change with Machine Learning workshop, which aims to demonstrate that AI can be an invaluable tool in helping society adapt to and mitigate the effects of climate change.

\subsection*{Ethics}
Inherently characterized by ongoing perplexity, ethics aspires for certainty and consensus, yet also remains dynamic and evolving. It has its origins in the work of philosophers such as Aristotle, who posited that ethics connects theory with \textit{praxis}, with the goal of guiding human actions towards \textit{eudaimonia} (i.e. the highest human good)~\cite{aristotle350}. In the millennia since Aristotle, the philosophical sub-domain of applied ethics has sought to establish normative principles for a variety of human activities and domains, which inexorably depend on the context of application and the individuals involved, leading to much debate regarding which norms should be applicable in which contexts, as well as the definition of key ethical concepts such as  fairness~\cite{saxena2019how,hutchinsonmitchell2019}, transparency~\cite{eyert2023,norval2022} and, indeed, the very definition of ethics itself ~\cite{bietti2020ethics,terzis2020onward}. 

Lacking consensus, the field of AI ethics often applies Western moral theories ranging from utilitarianism~\cite{mill1863, bentham1789} to egalitarianism~\cite{doi:10.1073/pnas.2213709120} and virtue ethics~\cite{aristotle350, aquinas1702} to propose ways of assessing the ethicality of AI systems. However, the application of these moral theories faces challenges given the difficulty of, e.g. quantifying the concept of utility in utilitarianism, assessing and comparing who is worse off in egalitarianism, or evaluating cultural variability in defining virtues in virtue ethics. This is also the case in terms of the application of these theories in modern-day contexts involving new types of AI-driven technologies such as robots or autonomous vehicles, which can be limited without a comprehensive understanding of both AI's technical capabilities (e.g. the limitations of the underlying models) as well as the diversity of life experiences of the people using these tools, who can interact with them in ways that are hard to predict or design for~\cite{lacroix2022metaethical}. Given that these concepts encompass work from a multitude of domains that espouse different objectives, values and methods~\cite{birhane2022forgotten}, their definition can have major consequences on the way in which these concepts are operationalized in AI models and systems~\cite{selbst2019fairness,friedler2021possibility,ashurst2022disentangling}. Work that addresses the ethical aspects of AI systems is discussed and published in conferences such as the ACM Conference on Fairness, Accountability, and Transparency (FAccT), as well as the AAAI/ACM Conference on AI, Ethics, and Society (AIES), which both have a cross-disciplinary focus and cover a multitude of topics in terms of the societal and ethical aspects of AI.

\section{Existing Scholarship in AI Ethics and Sustainability} \label{sec:theory-practice}

In the sections below, we explore existing scholarship in order to critically analyze how both ethics and sustainability are defined in theory and operationalized in practice within the diverse communities that pursue AI. First, we describe the principles and frameworks that have been proposed to guide AI from both an ethical and environmental perspective; next, we examine how these principles are applied, implicitly and explicitly, in AI research and practice. Finally, we look at several recent approaches for regulating and governing AI and the different roles adopted by stakeholders and organizations. 

\subsection{Principles and Frameworks} \label{subsec:principles}  

A common starting point to ensure the ethical development and deployment of AI systems is the definition of a structure for guiding this process, which can be operationalized via sets of principles or a framework. On the one hand, ethical principles are often defined at a high level, describing values and concepts outside of any specific context of deployment. As such, multiple sets of guiding principles for `ethical AI' have been proposed by different organizations ranging from research institutes to nonprofit and for-profit entities. However, given the many different types of AI approaches that exist, as well as contextual factors that influence their application, it is difficult to define universal, or even generalizable, guidelines. To this point, a 2019 analysis by Jobin et al. reviewed 84 sets of guidelines
mentioning a variety of principles, finding very limited convergence between them but identifying the values of transparency, fairness, non-maleficence, privacy and responsibility as being most common ~\citeyearpar{jobin2019global}.  Similarly, a multitude of ethical frameworks have been proposed, with most of them espousing specific visions of ethics by putting an emphasis on aspects ranging from human empowerment~\cite{floridi2018ai4people} to virtue ethics~\cite{hagendorff2022virtue}. In comparison to principles, ethical frameworks often aim to frame ethics from the perspective of implementation, identifying how challenges can be addressed and how to build consensus around ethical values, often anchored to specific contexts of deployment of AI systems, such as medicine~\cite{mccradden2023s} or autonomous vehicles~\cite{leikas2019ethical}. However, the technological mechanisms proposed to operationalize these frameworks are often defined in abstract terms, which have been found to be difficult to implement in practice from an engineering perspective, making them difficult to operationalize in practice ~\cite{prem2023ethical}. In terms of environmental sustainability, the UN SDGs are most commonly used to inform AI frameworks~\cite{gill2022conceptual}, with inspiration from fields such as ecology to guide the definition of methods based on metrics and evaluation methods that enable a more holistic assessment of AI's environmental impacts.  ~\cite{liao2022sustainability}. However, similarly to AI principles, there is an equal multiplicity of AI frameworks, and multiple analyses have been carried out with the goal of establishing overlap and transversal connections, which were found to be lacking~\cite{prem2023ethical,vesnic2020societal,barletta2023rapid}. 

Conversely, analyses of ethical AI principles have also observed a general lack of recognition of AI's environmental impacts within the different sets of principles that have been defined. Different reasons have been proposed for this lack of connection, from the reliance of most AI principles on traditional Western ethical perspectives, which are human-centered and assign intrinsic value to human beings above other living things~\cite{stanford-ethics-environment}, to the paucity of research on AI’s environmental impacts, which hinders the development of coherent principles~\cite{bolte2022from}. When sustainability is addressed in ethical AI frameworks, it is once again only limited to environmental sustainability, notably carbon footprint estimation. For instance, several industry-led sets of principles specifically addressing the environmental sustainability of AI have been proposed in recent years by organizations such as Salesforce~\citeyearpar{salesforce2024} and the Green Software Foundation~\citeyearpar{gsf2023}. While these propositions touch upon metrics such as energy efficiency, they do not address topics such as rebound effects~\cite{luccioni2025efficiency}~\footnote{The relationship between efficiency and sustainability is far from straightforward, given phenomena such as rebound effects, in which improved efficiency of a given technology can lead to increased usage of it and therefore increase the overall consumption of resources -- see~\cite{borza2014connection, luccioni2025efficiency} for a more in-depth review.}, transparency and access to compute, which we consider to be core to these discussions and which we discuss in Section \ref{sec:links}. Nonetheless, certain frameworks, such as the UNESCO recommendations on the ethics of artificial intelligence~\citeyearpar{unesco2021recommendation} do emphasize the importance of sustainability and of evaluating technologies based on their environmental impacts via the UN SDGs, which, as discussed in the introduction, have sustainable development at their core -- which we explore in more depth in subsequent sections.

\subsection{Research and Practice} \label{subsec:research} 

Given that AI is a distributed field consisting of a multitude of practitioners and organizations, the practical application of the principles and frameworks described in the previous section can differ immensely. In a 2022 study of papers submitted to conferences such as ICML and NeurIPS, Birhane et al. analyzed the values that were highlighted by their authors -- i.e. the positive attributes of their project that they emphasized and the negative impacts they considered explicitly~\citeyearpar{birhane2022values}. From the 59 values they identified, the most emphasis was put on aspects such as technical progress, quantitative evidence, and novelty, whereas ethical considerations around values such as beneficence, interpretability and respect for privacy (which are core to many AI principles and frameworks cited above)  were present only in a fraction of papers. Also, not a single one of the values Birhane et al. identified was explicitly linked to environmental sustainability, highlighting once again the lack of connection in the research community with sustainability writ large. 

Progressing from this observation, while sustainability-oriented research has not been prominently featured in venues that directly address AI ethics, it remains a topic of research that has been gathering momentum in recent years. The first research to formally address the environmental impacts of training AI models was the seminal 2019 article by Strubell et al. which quantified the carbon footprint of training BERT, a large language model (LLM), as reaching 626,155 pounds of $CO_2$ emissions~\cite{strubell2019energy}. Follow-up work by other researchers has shed more light on this issue, revealing different aspects of model architecture~\cite{patterson2021carbon} and training procedure~\cite{dodge2022measuring} that can impact its ensuing carbon footprint. The first proposal for “Green AI”, i.e. AI research that takes environmental impacts into consideration when training AI models, was made by Schwartz et al. in 2020~\cite{Schwartz2020green} - it was subsequently broadened to include aspects such as hardware and scaling~\cite{wu2021sustainable} and model deployment~\cite{luccioni2022estimating} more recently. However, this field of research, while increasingly prolific, has failed to take ethical considerations into account in its analyses, focusing solely on aspects such as carbon intensity and energy efficiency, and not on issues such as the environmental impacts of increased consumption due to the use of AI in targeted advertising~\cite{kaack2022aligning}, or the application of AI in the oil and gas sector~\cite{oilinthecloud}, which are liable to counter-balance any actual efficiency gains~\footnote{Recent media coverage of Microsoft's sustainability promises has estimated that a single contract to use AI to expand oil production ``could enable carbon emissions adding up to 640 percent of the company’s carbon removal targets"~\cite{grist2024}.}.
 
\subsection{Governance and Regulation} \label{subsection:regulation} 
Governance and regulation aim to establish mechanisms for decision-making, guiding the development and deployment of AI systems, and outlining the roles and responsibilities of each party involved~\cite{zwikael2015project}. 
There are many distributed efforts for governance in AI whose aim is to develop ethical guardrails, with some highlighting the importance of international institutions~\cite{ho2023international}, and others focusing on the public sector~\cite{kuziemski2020ai} -- reflecting that both bottom-up and top-down endeavors are useful to establish functional mechanisms for governing AI. If we take a look at recent community endeavors for AI governance, the 2022 Big Science workshop proposed a bottom-up approach that established mechanisms for various ethical aspects of the project such as data governance, quality metrics, and fostering stakeholder collaboration and transparency~\cite{jernite2022data}, as well as drafting  a consensus-driven ethical framework for governing the resulting artifacts that encompasses both legal and technical dimensions~\cite{strongertogether}. Interestingly, Big Science was one of the few projects that also considered and documented the carbon footprint of model training, evaluation and deployment, proposing a holistic, life cycle approach to estimating emissions~\cite{luccioni2022estimating}. There have also been proposals arguing for putting sustainability in the center of AI development and deployment~\cite{van2021sustainable}, as well as frameworks for certifying the sustainability of AI systems~\cite{bolte2022from}, across all the different pillars of sustainability (i.e. social, environmental and economic)~\cite{genovesi2022acknowledging}. However, we are still at the beginning of building governance mechanisms that offer a more comprehensive analysis of the impacts of AI systems from the perspective of ethics and sustainability.

A pivotal example of top-down governance is the European Union's AI Act, which draws from broadly defined ethical principles to inform its regulations~\cite{AIACT2022}. In fact, the text of the Act demonstrates considerable progress following the recent European dialogues, seeking to regulate AI applications that may infringe on human rights, adhering, among others, to the ethical principles of human oversight, human agency and transparency. Moreover, the AI Act’s foundation on the premise that risk equates to potential human rights harms~\cite{act2023deal} echoes one of the longstanding traditions in AI ethics of prioritizing human rights \cite{aiethicsrights, ashrafian2015intelligent}. This approach embodies the ethical commitment to safeguard fundamental freedoms and human rights in the digital era, ensuring that AI technologies do not infringe upon these important principles. However, while both EU AI Act~\citeyearpar{AIACT2022}, as well as similar regulatory initiatives in  China~\citeyearpar{chineseai2023} and Canada~\citeyearpar{aida2024} point to the need to protect both fundamental human rights as well as to limit damage to environment, there are no official provisions regarding sustainability in any of their texts, and it remains to be seen how existing standards for environmental impacts in all of these jurisdictions will apply to AI systems. Similarly, sustainability considerations were also lacking in the 2023 US Executive Order regarding AI~\cite{biden2023executive}, which did not mention AI's greenhouse gas emissions nor energy usage, as well as multi-nation declarations such as the {Bletchley Declaration~\citeyearpar{bletchley2023}, illustrating the disconnect between sustainability and ethics in recent approaches to AI regulation. 

\section{Transversal Issues in AI Ethics and Sustainability} \label{sec:links}

\begin{displayquote}
\textit{"Is it fair … that the residents of the Maldives (likely to be underwater by 2100) or the 800,000 people in Sudan affected by drastic floods pay the environmental price of training and deploying ever larger English LMs, when similar large-scale models aren't being produced for Dhivehi or Sudanese Arabic?}" 

\flushright{Bender, Gebru et al.~\citeyearpar{bender2021dangers}}
\end{displayquote}

In the current section, we define four recurring issues that we have found to be particularly salient to discussions around both AI ethics and sustainability. These issues are inspired by previous carried out by critical scholars such as Dobbe and Whittaker~\citeyearpar{dobbe2019ai}, Birhane~\citeyearpar{birhane2022forgotten}, van Wynsberghe~\citeyearpar{van2021sustainable} as well as Bender and Gebru~\citeyearpar{bender2021dangers}, as cited above. We start, in Section~\ref{subsection:bias} with a discussion of the perils of assumptions of the generalizability of data and models in both ethics and sustainability. We follow, in Section~\ref{subsection:evaluation} with a reflection upon the current state of evaluation of AI systems, what is measured, what is missing, and why that matters. Next, we remark upon the current lack of transparency with regards to information relevant to both ethics and sustainabilityin Section~\ref{subsection:transparency}. Finally, in Section~\ref{subsection:power}, we discuss the balance of power and the allocation of justice, and how existing inequalities can be further amplified by AI systems.

\subsection{Generalizability} \label{subsection:bias} 

AI technologies function based on assumptions of generalizability and representativeness -- i.e. that given sufficient data, an AI model can learn to accurately represent (any) given process and even adapt to previously unseen data~\cite{goodfellow2016deep}. For instance, the concept of pre-training AI models on large datasets such as ImageNet~\cite{deng2009imagenet} dates back to the early 1990s~\cite{Schmidhuber1991neural} and has since become the dominant training paradigm in both computer vision~\cite{szegedy2015going,redmon2016look} and natural language processing~\cite{devlin2018bert,liu2019roberta}. In fact, pre-training is heavily dependent upon the assumption of representativeness - i.e. that the huge amounts of training data used for pre-training represent the world as a whole, or at least a sufficient part of it to be useful for any kind of downstream application (i.e. fine-tuning, transfer learning, etc). While the limitations of such claims for both AI models and datasets have been previously shown (see ~\citet{smith2022real,koch2021reduced,raji2021ai,chasalow2021representativeness}), the theory of generalizability, and the perception of certain types of AI models, ie. LLMs, as ``general purpose technologies'' continues to persist~\cite{eloundou2023gpts}. This can come with both ethical and environmental ramifications when systems trained under the pretense of generalizability are applied in contexts that differ from the ones represented in their training data -- we discuss some of these below.

Given the data that fuels AI models is produced by humans, it is intrinsically laden with subjective judgments~\cite{miceli2022studying} and representative of specific worldviews~\cite{davis2020dataset}. Numerous studies have shown that both the data used for training AI models ~\cite{dodge2021documenting,rogers2021changing,raji2021ai} and the models themselves~\cite{gonen2019lipstick,bender2021dangers, wolfe2022american} are not, in fact, representative of the world at large and that the biases contained in training data persist even if further fine-tuning is carried out~\cite{ladhak-etal-2023-pre}, which can have `cascade' effects when models are deployed in production~\cite{sambasivan2021everyone}, which can contribute to perpetuating negative biases~\cite{gebru2019oxford}. Similarly, off-the-shelf, proprietary technologies that are marketed as generic can fail at the tasks when applied in differing contexts, e.g. in applications such as facial recognition (when they fail to recognize people from populations under-represented in training datasets~\cite{buolamwini2018gender}) and crime prediction (where they have dismal accuracy rates across different locations~\cite{predictivepolicing2023})
-- and yet, out-of-the-box AI systems for these tasks and many others continue to be built and deployed under the assumptions that they will work no matter the context of their application. This can have devastating effects on already marginalized communities when applied for tasks such as criminal sentencing~\cite{angwin2022machine}, and facial recognition~\cite{buolamwini2018gender}.

Similarly to data concerning human beings, environmental and ecological data can also contain biases, for instance  in its temporal coverage and geographical spread, which can be as damaging to biomes~\footnote{A biome is a bio-geographical unit that corresponds to a community of plants and animals that share a physical environment and a climate.} of plants and animals as they are to communities of humans~\cite{joppa2016filling,siddig2019biodiversity}. From a modeling perspective, AI models trained to carry out biodiversity monitoring on one ecosystem often fail to perform accurately on others, no matter the species~\cite{sethi2023limits,koh2021wilds}, and yet ecological bias assessments are not often carried out before model training. In fact, White et al. refer to geographical differences in ecological datasets as the main factor limiting our capacity to predict how biodiversity will be impacted by future changes in the climate, notably due to missing data from regions such as Africa and South America, making data-driven approaches such as AI unrepresentative of entire continents~\cite{white2021geographical}. 

An example of the dire consequences this can have can be found in the field of short-term climate modeling and disaster detection, which relies on data from sensors and weather radar stations that allow for events such as floods and wildfires to be tracked in real-time across borders, oceans and continents. 
While geospatial data is inherently global (given that satellites circle the planet as a whole), it relies upon assumptions of generalizability that often fail to hold when applied in contexts that differ from the training data, or the fact that many regions are missing these data sources for reasons ranging from insufficient funding and aging technological infrastructure to lack of connectivity~\cite{chason2021lack,tzachor2023reduce}, which results in data gaps~\cite{sefala2021constructing, tseng2021cropharvest}. This can come at a cost to living beings, both human and animal, in the regions and contexts where these technologies are applied -- for instance in early warning systems for extreme weather events~\cite{otto2023without}.and deforestation detection~\cite{kinnebrew2022biases}. Of course, any assessment of representativity (or lack thereof) hinges upon an appropriate evaluation of this assumption -- which we address in the following section. 

\subsection{Evaluation} \label{subsection:evaluation}

A popular adage states that “you can't improve what you don't measure”~\footnote{The quote is often attributed to Peter Drucker, although its exact origins are unclear.}; in the context of AI systems, this can be translated into the fact that the criteria that we use to evaluate AI systems and the way in which this evaluation are carried out are important -- i.e. the metrics we choose help us embed our values as communities about outcomes we wish to see and those that we put less emphasis on~\cite{mitchell2020diversity}. While leaderboards such as Papers With Code tend to only measure performance-based metrics such as accuracy or precision, factoring in other metrics can make comparisons between different models more meaningful and actionable. This is due to the fact that real-world constraints on model deployment often result in trade-offs being made between different factors that include accuracy and efficiency~\cite{brownlee2021exploring}, but also robustness~\cite{zhang2019theoretically}, inclusion ~\cite{hooker2020characterising} and data quality~\cite{baeza2017quality}. This means that in order to meaningfully assess the utility of AI systems in different practical contexts, other measures must be considered in parallel to performance and accuracy~\cite{lucivero2020big}; and often the best people to do the assessments for the trade-offs are the populations who will use these tools. For example, when it comes to the evaluation of generative technologies such as large language models, these do not have a single well-established evaluation approach~\cite{raji2020closing,mokander2023auditing}. Approaches that are used for evaluating their ethical limitations include red-teaming~\cite{ganguli2022red}, external audits~\cite{mokander2023auditing,raji2023change} as well as more holistic model evaluations that reflect different aspects of model performance~\cite{bommasani2023holistic,eval-harness}. However, critiques of the approaches described above also include their non-inclusivity of marginalized communities~\cite{dennler2023bound} as well as a lack of formalized approaches and standards for model evaluation, making apples-to-apples comparisons between models difficult~\cite{costanza2022audits}. Also, as many of the most widely deployed AI systems are currently proprietary and direct access to models is not possible, it is hard to exhaustively evaluate many popular commercial models for any meaningful evaluation to take place~\cite{solaiman2023evaluating, luccioni2023mind}.

Similarly, evaluating the environmental impacts of AI systems is far from straightforward, and we are still missing many pieces of the puzzle needed in order to meaningfully estimate these impacts. For instance, most of the carbon footprint assessments only focus on the training stage of AI models, which is easier to quantify and report~\cite{strubell2019energy,patterson2021carbon}, but which only represents a portion of models' total environmental impacts. In a 2023 article estimating the carbon footprint of BLOOM, a 176 billion parameter LLM, Luccioni et al. proposed using a Life Cycle Assessment approach for this evaluation, since it takes into account different stages of the model life cycle including the manufacturing of computing hardware, idle energy usage, and model deployment, finding that training accounted for only half of the model's overall emissions~\cite{luccioni2022estimating}, meaning that similar studies that only took training into account were potentially underestimating their emissions by half. Also, while commendable in terms of its granularity, this kind of carbon accounting fails to recognize the societal and economic aspects of sustainability, such as the contribution of LLMs such as BLOOM towards amplifying the existing inequalities in the field of AI due to the increased amount of computing resources that they require, which are unattainable to many members of the AI community, as well as the propagation of biases via their usage. Furthermore, the authors themselves note that there is currently no information available about the embodied emissions linked to manufacturing GPUs, so it is impossible to estimate what portion of the overall carbon footprint this represents. This highlights that the emphasis on environmental sustainability often fails to account for other aspects of AI's global impacts -- and any kind of evaluation hinges upon transparency, which is sorely lacking in the field of AI -- we discuss this in more length in the following section. 

\subsection{Transparency} \label{subsection:transparency}

Transparency is widely recognized as a fundamental principle in science in general and AI in particular~\cite{felzmann2020transparency, larsson2020transparency, wischmeyer2020AItransparency, walmsley2021AItransparency} but actualizing it in practice can be challenging. This is, in part, due to the fact that machine learning-based systems are not inherently transparent or interpretable, given the complexity of the neural network architectures they espouse and the number of parameters they contain~\cite{molnar2020interpretable,lakkaraju2019faithful,lipton2018mythos}. Efforts such as interpretability approaches are useful and can help interpret the predictions of models posthoc~\cite{ribeiro2016model,lakkaraju2016interpretable}, whereas artifacts such as data sheets and model cards~\cite{mitchell2019model,gebru2021datasheets} can contribute towards making AI systems more understandable for users, providing essential information about AI models in a user-friendly format. These artifacts  allow users to understand not just how an AI system functions, but also its limitations, potential biases, implications, and environmental impacts.
However, even though model cards are increasingly used in practice (for instance by AI model-sharing platforms such as Hugging Face) and provide important information about models, they are not sufficient to guarantee, for instance, the reproducibility of reported results, which is a core tenet of scientific practice~\cite{national2019reproducibility}. 

Indeed, several studies of transparency found that an overwhelming amount of results published at technical AI conferences do not document all of the variables necessary to reproduce the results they report~\cite{gundersen2018state,raff2019step}. This situation highlights the need for an approach to transparency that would involve reporting but also ensuring the reproducibility of AI models and their findings. Such an approach would acknowledge the connection between transparency and reproducibility: transparent research practices enable reproducibility, which in turn facilitates independent scrutiny, validation, further development of research findings by other scientists \cite{haibe2020transparency}. The absence of transparency, especially in sharing essential materials such as model weights, code and data, impedes the ability to reproduce results, diminishing AI models' scientific impact and impeding their adoption within the wider scientific community~\cite{haibe2020transparency}.

In terms of sustainability, the AI community has historically been even less transparent regarding the environmental impacts of AI models and systems, with most work in this field being done post-hoc by researchers who did not do the initial model training and deployment (e.g.~\cite{strubell2019energy,luccioni2023counting}). The most common environmental sustainability metric for AI models, their carbon footprint, is rarely, if ever, disclosed.  While model cards of recent models such as BLOOM~\cite{workshop2022bloom} and Stable Diffusion~\cite{Rombach_2022_CVPR} have included carbon footprint information, it remains far from common information communicated by model creators -- recent work has found that the overwhelming majority of models shared publicly do not include this information~\cite{castano2023exploring}. In fact, most carbon footprint analyses gather the information manually by writing to authors. For instance, Luccioni and Hernandez-Garcia reached out to over 500 authors of AI papers to get information needed to estimate the carbon footprint of their models, and were only able to collect 95 answers, with many authors refusing to provide the relevant information, citing privacy concerns and lack of experimental logs~\citeyearpar{luccioni2023counting}. In fact, until recent years, the general emphasis in the AI community was put on the efficiency and `greenness` of AI as opposed to its environmental costs, which are now slowly starting to gain traction as an important consideration for AI systems~\cite{hogan2018big,monserrate2022cloud}. %Interestingly, ensuring the replicability of results also comes with an added benefit for sustainability, since it avoids the necessity of having to re-run experiments and empirical evaluations, therefore saving computational resources and wasted energy. 
In fact, given the increasing size and computational requirements of models being productionized in recent years (especially LLMs), training them is only accessible to a small fraction of the AI community, which means that the organizations who have the necessary resources for this have a disproportionate influence on the field as a whole--- we discuss this in the following section.

\subsection{Power and Equity} \label{subsection:power}

Modern AI research and practice are not equitable by design: their cost in terms of computer hardware as well as human skills means that only a small percentage of both academic and industrial organizations can contribute to many aspects of model development. With the recent advent of AI models of ever-increasing scale and complexity, the digital divide in AI is only increasing, as it takes more compute and human skill to train and deploy AI models and systems~\cite{chan2021limits,abdalla2021grey,luccioni2023mind,besiroglu2024compute}. This means that the future wide-sweeping benefits that AI technologies are promised to have for humanity as a whole~\cite{ordonez2023openai} are contingent upon access to technologies that are fundamentally unequally distributed. Despite explicit proposals to pursue more equitable and explicitly de-colonialist approaches~\cite{madianou2021nonhuman,mohamed2020decolonial}, the `bigger-is-better' paradigm continues to be central to AI research and practice~\cite{varoquaux2024hype}.
In a similar fashion, the places where AI research is being carried out are also skewed towards institutions from a handful of countries mostly located in the Global North~\cite{abdalla2023elephant}, which inexorably impacts the choices made during the AI development and deployment process, introducing many biases (which we have already addressed in previous sections).  In fact, recent work has proposed that the very pursuit of sustainable AI has the opposite effect, contributing towards maintaining the status quo and ``securing the dominant socio-economic interests of neo-liberal capitalism"~\cite{schutze2024problem}. For instance, major technological corporations have dedicated significant resources towards solutions such as improving the efficiency of their data centers, proposing numerous initiatives towards technological sustainability~\cite{aws, gcp}, including research at the nexus of AI and the climate~\cite{merchant2023scaling,bcg_google}. However, both Microsoft and Google announced that they would miss their 2024 sustainability targets, due in part to the energy demands of the AI tools that they have been developing and deploying~\cite{rathi2024microsoft,metz2024google}. The impacts of these computation-intensive data centers can further be expanded to include the mining of rare metals and the disposal of e-waste, both of which predominantly affect countries from the Global South which profiting technology companies from the Global North~\cite{taffel2019ecocide,hogan2018big}.

Similarly to AI, issues of power, equity and justice are also central in the field of sustainability, since climate change is an inherently inequitable phenomenon -- with a handful of countries and regions in North America, East Asia and Europe responsible for a disproportionate portion of global emissions, while the impacts of sea level rise and extreme weather events being felt most strongly in countries with very minimal carbon footprints, raising questions of equity and justice and how to address them~\cite{page2008distributing,schlosberg2014environmental,denton2002climate,mclaren2018whose}. Similarly, the majority of climate-focused AI solutions  overlook issues of justice and power, focusing predominantly on the climate-positive aspects of technologies and not who stands to benefit from them, or whether these issues can be solved with technology in the first place~\cite{brevini2023myths}.For instance, when  AI systems that carry out precision agriculture are developed, the emphasis is made on efficiency or increased crop yields~\cite{sharma2020machine}, and not the fact that these systems are liable to replace already underprivileged communities such as migrant workers that traditionally harvested crops by hand, or the disparate impacts of proposed solutions on different communities~\citep{waldmueller2015agriculture,durokifa2018neo,ziai2016development}. In addition, AI technologies can be seen as further exacerbating the existing inequalities in terms of the distribution of power and profits~\cite{heilinger2023beware} and perpetuating the existing extractionist approaches in terms of labor by exploiting already under-paid and marginalized workers for data collection and labeling tasks~\cite{ricaurte2022ethics} as well as creating a new class of precarious crowd-sourced workers~\cite{williams2022exploited}, which are often located in the Global South or in already marginalized communities, where the impacts of climate change have reduced the viability of traditional professions such as farming~\cite{rhee2018robotic,hao2022artificial}. This is an additional example for why we advocate for establishing more concerted efforts for integrating AI ethics and sustainability -- we describe these in the following section.

\section{Establishing Best Practices for AI Ethics and Sustainability} \label{sec:ways-forward}

Having established a multitude of transversal topics that span  AI ethics and sustainability, we now focus on proposing best practices that would integrate the two in AI research and practice. We draw upon existing work to show how we can build upon it both within the AI community and in tandem with members of other communities ranging from sustainable development to policy-making and engineering.

\begin{table}[h!]
\small
\begin{tabular}{l|l|l|l} & \textbf{Principles and Frameworks}    & \textbf{Research and Practice} & \textbf{Governance and Regulation}  \\ \hline
Generalizability & \begin{tabular}[c]{@{}l@{}}Shifting from the dominant \\ Western moral philosophies \\ to include perspectives from\\  non-Western traditions\end{tabular} & \begin{tabular}[c]{@{}l@{}}Studying how well AI models \\ generalize to different populations\\ of human and non-human living\\ beings\end{tabular}           & \begin{tabular}[c]{@{}l@{}}Espousing bottom-up governance \\ approaches based on the cultural, \\ societal and geographical constraints\\ of AI system deployment\end{tabular} \\ \hline
Evaluation       & \begin{tabular}[c]{@{}l@{}}Developing frameworks that \\ accommodate both ethical and\\ environmental responsibility\end{tabular}                          & \begin{tabular}[c]{@{}l@{}}Carrying out more holistic\\ evaluation of models and systems, \\ spanning both ethical and \\ environmental criteria\end{tabular} & \begin{tabular}[c]{@{}l@{}}Integrating both social and\\ environmental assessments \\ into existing and in-progress \\ AI regulation\end{tabular}                              \\ \hline
Transparency     & \begin{tabular}[c]{@{}l@{}}Broadening the scope of \\ transparency to include its\\ social and environmental aspects\end{tabular}                          & \begin{tabular}[c]{@{}l@{}}Communicate the costs \\ and impacts of AI systems\\ on both the environment \\ and society\end{tabular}                           & \begin{tabular}[c]{@{}l@{}}Requiring deployed AI systems\\ to carry out audits and report \\ a minimum of metrics spanning\\ both bias and ethics\end{tabular}                 \\ \hline
Power            & \begin{tabular}[c]{@{}l@{}}Developing principles and \\ frameworks that address\\ both human and ecological needs\end{tabular}                             & \begin{tabular}[c]{@{}l@{}}Make equity-informed \\ trade-offs when developing \\ and deploying AI\end{tabular}                                                & \begin{tabular}[c]{@{}l@{}}Involving multiple stakeholders, \\ especially ones from the concerned \\ communities and areas, in the \\ governance process\end{tabular}         
\end{tabular}
\caption{A summary of the proposed best practices for different axes that we cover in our article}
\end{table}
\vspace{-30pt}

\subsection{Principles}

In the context of AI, there are multiple facets of AI technologies that have to be taken into consideration, given the intersection between AI and the broader societal context in which it operates. Importantly, integrating broader sustainability into AI guidelines ensures that justice and fairness are not just about social dimensions but also include respecting and protecting the environment -- i.e. expanding the definition of sustainable AI to include the social and economic pillars proposed by Barbier~\cite{barbier1987concept}. This integration also acknowledges that true justice in AI cannot be achieved without considering its environmental and societal implications, which helps build the bridge between complementary approaches in both AI ethics and (environmental) sustainability research.

\paragraph{Generalizability} Given the observed disconnect between ethics and sustainability in the context of AI principles and frameworks, we find that, while these offer a valuable starting point, they often fall short in addressing AI's complex ethical issues due to their lack of contextual sensitivity~\cite{munn2022}. We also believe that improving upon this necessitates a more nuanced and context-specific approach to AI ethics, one that embraces the varied ethical dimensions presented by AI, including its environmental implications. Current ethical charters in AI, often detached from this perplexity, represent preliminary thoughts on the ethical landscape but lack the depth required for many practical applications~\cite{andler}. By recognizing these limitations, we intend not to discard these definitions and principles but to improve upon them. In this context, environmental and sustainability challenges related to AI development and deployment are integral to the broader ethical reflections within the field. This integration between ethics and sustainability in AI calls for a holistic approach, where ethical considerations are not viewed in isolation but are intrinsically linked with environmental and sustainability oversight. For instance, shifting from the dominant Western moral philosophies to include perspectives from non-Western traditions such as relational ethics~\cite{metz2016}, Ubuntu ethics~\cite{nagel2022}, and Confucian ethics~\cite{li2013} can offer valuable insights into community, social harmony, and the interconnectedness of beings, emphasizing the impact of AI on society and interpersonal relationships.

\paragraph{Evaluation} There is no one-size-fits-all solution for either ethics or sustainability and, indeed, no single way of concluding that an AI system is neither truly ethical nor sustainable. Recent work has begun bridging the gap; for instance, work by Lynch et al. is inspired by the concept of urgent governance in environmental studies, which distinguishes system reliability and societal harm and advocates for the consideration of both when auditing infrastructure and technologies~\cite{lynch2018urgency}. Raji et al. use a similar approach for their proposed end-to-end framework for internal algorithmic auditing of AI models, which takes into account both technical and ethical assessments~\cite{raji2020closing}. In a similar vein, a recent model evaluation framework by Rakova et al., proposes an environmental justice-oriented lens to carry out algorithmic audits~\cite{rakova2023},  that of Genovesi and Mönig places sustainability at the center of Ethical AI certification~\cite{genovesi2022acknowledging}, while that of Metcalf et al. uses environmental impact assessments as an example of a formal mechanism that can be used to inspire the assessment of AI technologies~\cite{metcalf2021algorithmic}. All of these approaches acknowledge that ethical decisions in AI have environmental consequences and vice versa, thus necessitating a framing that accommodates both ethical and environmental responsibility.

\paragraph{Transparency} When viewed as a means to foster greater accountability, transparency takes on a central role: it becomes a principle that enhances ethical compliance and promotes environmental responsibility, contributing to sustainability. For instance, integrating social transparency and sustainability can be exemplified by an AI system designed for urban planning, such as an AI tool developed to optimize city layouts for efficiency. 
In this example, to include sustainability approaches,  developers would also provide information on the environmental footprint of running the AI system, such as energy consumption during data processing and potential environmental benefits of the proposed urban layouts, like reduced carbon emissions from optimized traffic flows or green spaces. In this way, by deepening the concept of transparency to include social and environmental aspects, we would move towards creating AI systems that are more robust, socially responsible and ultimately more accountable about the environmental impacts they have and making more informed decisions based on the information at our disposal~\cite{diakopoulos2020accountability}.

\paragraph{Equity and Power} Equity is about ensuring fair access and participation in the benefits and governance of technology across different communities, especially those historically marginalized. Philosophical perspectives on equity, drawing from theories of distributive justice~\cite{lamont2017distributive, miller1995pluralism}, emphasize the necessity of equitable distribution of resources and responsibilities and risks associated with AI technologies.
This principle is especially relevant in environmental justice, where the disproportionate impact of environmental harms on specific populations demands a reevaluation of AI technologies are deployed at scale -- for instance, Schlosberg's theory of recognitional justice highlights the importance of recognizing and respecting diverse community needs and values in environmental policies ~\cite{schlosberg2012climate}.
Additionally, equity requires that AI development actively includes diverse voices in its creation and implementation phases, ensuring that AI systems do not perpetuate existing disparities but rather contribute to rectifying them. This approach draws upon environmental justice literature for developing principles and frameworks that addresses both human and ecological needs, thus framing equity as a matter of distribution, procedural and interactional fairness~\cite{bullard1997confronting}.

\subsection{Research} Despite a lack of common terminology, similar issues arise both in terms of considerations of AI ethics and sustainability and considering the inter-connectedness of the two when designing and deploying AI systems is paramount given their socio-technicality and the consequences this can have on the human and non-human species residing in these regions. 

\paragraph{Generalizability} At the nexus of AI ethics and sustainability, existing scholarship has already established that the most disadvantaged and marginalized members of our societies tend to be the least well-represented in the `Big Data' used in many AI models and systems~\cite{cullen2001addressing,shenglin2017digital,gebru2019oxford, buolamwini2018gender}; the same applies at the level of countries and regions, with `global' datasets reflecting things like wealth and economic development only being tested in a select few countries ~\cite{blumenstock2018don,hilbert2016big} and the most extensive biodiversity datasets consisting of data collected in a subset of regions from the Global North, as well as regions close to cities and roads~\cite{beck2014spatial,troudet2017taxonomic}.  Studying the limits of application of AI systems both in terms of ethics and sustainability and how well they generalize to different populations of human and non-human subjects is important to question existing assumptions. For instance, studies of emblematic datasets such as ImageNet found that it to misrepresent both humans~\citeyearpar{crawford2021excavating} and other living beings such as insects and fish~\citeyearpar{luccioni2023bugs}. Developing new datasets that are more representative of diverse populations and contexts - such as the Dollar Street dataset~\cite{rojas2022dollar} and CropHarvest~\cite{tseng2021cropharvest} - and using this dataset in research and practice and help improve the applicability of AI systems and their ability to represent more diverse populations from both a societal and environmental perspective.

\paragraph{Evaluation} As AI is increasingly used in the fight against climate change, holistic evaluation of models and systems becomes ever more relevant. For instance, in the high-stakes domain of solar geoengineering, which aims to develop new ways for modifying the Earth's climate to reduce the global warming effect (i.e. by enhancing the reflexivity of clouds so that they reflect more of the sun's rays), AI is often used to help model the potential far-reaching effects of even minor interventions and understand how they will impact local and global climate patterns~\cite{schroeder2019stratospheric,nowack2018using, feinberg2022solar}. This is because regional changes to the climate may trigger increased fragility of some regions and not others, which is hard to quantify and therefore, to be optimized for in AI models that aim to predict the consequences of climate interventions~\cite{heilinger2023beware}. For instance, different thresholds of solar reflexivity are optimal for different regions, and optimizing results based on a given region (e.g. the continental United States) would make things worse for others (e.g. Western Africa), which can suffer droughts and other forms of damage~\cite{mclaren2018whose}. Considerations around the wider rebound effects of proposed AI solutions are also important: for instance, in the case of AI systems that improve aircraft efficiency~\cite{le2023improving}, more efficient aircraft can result in cheaper airfare and therefore, more travel overall\footnote{This is often referred to as \textit{Jevons paradox}, which observes that when technological progress improves the efficiency of technology, this actually results in its increased usage and increases overall resource use~\cite{jevons2018coal}.}.

\paragraph{Transparency} Falk and van Wynsberghe argue that transparency is a crucial factor for establishing \textit{``whether the net environmental impact of AI for Sustainability is positive or how the positive impact on the respective sustainability goal can outweigh the very different negative impact of the models’ development on sustainability"}~\cite{falk2023challenging} -- we would expand this to include AI systems in general, which should use a variety of mechanisms to communicate the costs and potential impacts of their systems on both the environment and society. In fact, as proposed by Ehsan et al., the notion of transparency in AI can be expanded to encompass "social transparency", which involves integrating socio-technical aspects in the description and understanding of AI systems~\cite{socialtransparency}. Social transparency involves a portrayal of an AI system's societal impacts, ethical considerations, and eventually its environmental footprint. By doing so, it provides a more complete picture, making the AI system transparent but also understandable in a broader societal context. This augmented view of transparency, which would integrate both social and environmental dimensions, resonates with the increasing awareness that AI is not simply a technological tool but a socio-technical system with extensive repercussions, spanning both people and the environment~\cite{vandePoel2020}. 

\paragraph{Equity} Recognizing that Green AI (or sustainable AI), i.e. the development and deployment of AI systems that puts the emphasis solely on efficiency or the reduction of greenhouse gas emissions, is not necessarily inherently more equitable or just -- for instance, if the more efficient models are not widely shared, or entail an increased usage of compute due to their efficiency -- is an important first step towards improving the current direction towards bigger models. Recent research in both AI ethics and sustainability has shed light on the extent to which AI systems enable the amplification of existing social inequities~\cite{eubanks2018automating, benjamin2023race, lloyd2018bias} as well as contributing to the preservation of the ecological status quo vis-a-vis to climate change~\cite{taffel2019ecocide}. Given these findings, it is important to make equity-informed trade-offs when developing and deploying AI -- for instance by carrying out energy prediction in resource-constrained energy grids, which are common in low- and middle-income countries~\cite{bariyatopology2023}, or by using AI to predict the impacts of changes in climate on societal aspects such as disease propagation and health~\cite{kuehnert2022surrogate}. 

\subsection{Governance}

As the field of AI ethics increasingly intersects with regulation, including law and policy, it showcases its interdisciplinary nature, meaning that in order to be successful, governance initiatives must incorporate efforts from different domains, depending on the context of the application.

\paragraph{Generalizability} While there is no single solution to complex questions involving governance over AI systems, various bottom-up governance approaches have been proposed based on the cultural, societal and geographical constraints of AI system deployment. Some of these follow the tenets of the Māori culture, which is based on principles that consider both impacts on nature as well as on fellow human beings, bridging the gap between ethics and sustainability~\cite{munn2023five,hao2022new}; others espouse those established by the Indigenous AI community, which is based on both values and practices of social and environmental sustainability, both core to many Indigenous epistemologies~\cite{lewis2020indigenous}. Also, regulating the deployment of out-of-the-box AI solutions that operate on the premise of generalizability without taking context into account can help ensure that systems that are meant to be widely applicable are truly representative of the context of application -- where evaluations, as explained below, will play a crucial role.  

\paragraph{Evaluation} As noted by Metcalf et al.~\citeyearpar{metcalf2021algorithmic}, there is a parallel between environmental impact assessments and AI ethics audits, which can be extended beyond ethical compliance to include assessments of environmental impacts, such as energy consumption and carbon emissions. Requiring audits of commercial AI systems before their deployment in practice, both in contexts such as education and healthcare that come with high stakes in terms of societal impacts, but also in contexts such as disaster prediction and climate modeling, that come with potentially widespread environmental impacts, will require the development of new governance approaches. For instance, attempting to evaluate the wider rebound effects of AI tools and their impacts on consumption and human behavior is important to represent their broader impacts on both society and the environment~\cite{taffel2019ecocide,hogan2018big, kaack2022aligning}. Finally, integrating both social and environmental assessments into existing and in-progress regulation and developing new approaches to evaluate these impacts can help ensure that the deployment of AI systems is carried out in a way that is ethically sound and sustainable across multiple dimensions. 

\paragraph{Transparency} Recent years have seen less transparency in AI research and practice, especially in terms of generative AI models~\cite{solaiman2023gradient}. However, as these systems are increasingly being deployed in society, having more information regarding how these systems were created and deployed remains paramount. Ensuring that enough details are provided both regarding the energy consumed and greenhouse gasses emitted during model training and deployment can help track how the environmental impacts of AI are evolving over time. Mandating transparency for already deployed AI systems can help establish audits, red teaming efforts and AI energy score ratings to raise users' awareness around the impacts of the systems they use~\cite{brevini2020black}, contributing to what can be termed "usable transparency"~\cite{murmann2017usable}. By ensuring that the processes and results of these practices are well-documented and publicly accessible, stakeholders would be better equipped to understand and evaluate the ethical and sustainable aspects of AI systems. Such policies would promote a culture of openness in the AI industry, encouraging developers to prioritize ethical considerations and sustainability alongside technical advancements.  

\paragraph{Equity} Involving multiple stakeholders, especially ones from the concerned communities and domains, in this process is important to ensure that different perspectives and lived experiences are reflected in the development and deployment of AI systems~\cite{rotz2019automated, ryan2022social} as well as the different communities and can influence existing and future practices~\cite{mclaren2018whose,dara2022recommendations}. From a regulatory perspective, the Finnish ETAIROS (Ethical AI for the Governance of the Society) project proposed the integration of ethics, sustainability, design and foresight for inter-disciplinary governance of AI systems~\cite{mika2019ethical}, whereas the White House Office of Management and Budget (OMB)'s first government-wide policy around the usage of AI includes, \textit{inter alia,} clauses that stipulate that government agencies should take both environmental impacts and bias and fairness into account when procuring AI-enabled services~\cite{OMB2024} - exhibiting thought leadership that will hopefully have wider repercussions.

\section{Conclusion} \label{sec:conclusion}

We recognize that issues of ethics and sustainability are complex and, especially in the context of emerging technologies like AI, it can be difficult to define what progress looks like and how it can be achieved. We do not pretend to have developed a universal approach for either of these issues (and do not believe that one can exist) -- but by adopting a multitude of endeavors such as the ones described in the paragraphs above can help involve different actors and hopefully build momentum across the AI community. 
The beauty of making transversal connections that go above and beyond the silos in which many AI technologists tend to operate is that we can also be inspired by the multitude of rich and relevant work that has been done in other domains -- from ecology to philosophy, as well as governance and climate science, to propose ways forward that would allow the AI community to improve systems from the perspective of both ethics and sustainability. Looking forward, the field of AI ethics is rapidly evolving, with insights racing to keep up with the rapid pace of technological advancements in AI. This dynamic landscape presents an ongoing challenge: to develop AI in a way that is inclusive, just, and cognizant of its environmental and societal impacts. Furthermore, it is becoming increasingly clear that AI ethics and sustainability are interdependent: they must go hand in hand to ensure a holistic societal impact. The absence of either aspect leads to an incomplete perspective, potentially overlooking critical societal and environmental consequences. Therefore, integrating AI ethics with sustainability is not just beneficial but necessary, ensuring that AI advancements are not only technologically innovative and ethically sound but also maximizing their potential to engender sustainable advancement.

\clearpage

\bibliographystyle{ACM-Reference-Format}
\bibliography{references}

\end{document}